\documentclass[twoside]{article}
\usepackage{qic,epsfig}

\def\gtsim{\mathrel{ \rlap{\raise.5ex\hbox{$>$}}
                      {\lower.5ex\hbox{$\sim$}}  } }

\begin{document}
\bibliographystyle{plain}

\setlength{\textheight}{8.0truein}    

\runninghead{Surface-electrode architecture for ion-trap quantum
information processing}
            {J.~Chiaverini, R.~B.~Blakestad, J.~Britton, J.~D.~Jost, C.~Langer, D.~Leibfried, R.~Ozeri, and D.~J.~Wineland}

\normalsize\textlineskip \thispagestyle{empty}
\setcounter{page}{419}

\copyrightheading{5}{6}{2005}{419--439}

\vspace*{0.88truein}

\alphfootnote

\fpage{419}

\centerline{\bf
SURFACE-ELECTRODE ARCHITECTURE FOR ION-TRAP} \vspace*{0.035truein}
\centerline{\bf QUANTUM INFORMATION PROCESSING} \vspace*{0.37truein}
\centerline{\footnotesize
J. CHIAVERINI\footnote{email: john.chiaverini@boulder.nist.gov},
R.~B.~BLAKESTAD, J.~BRITTON, J.~D.~JOST, C.~LANGER,}
\centerline{\footnotesize D.~LEIBFRIED, R.~OZERI, and
D.~J.~WINELAND} \vspace*{0.015truein} \centerline{\footnotesize\it
NIST, Time and Frequency Division, Ion Storage Group}
\baselineskip=10pt \centerline{\footnotesize\it 325~Broadway,
Boulder, CO 80305 USA} \vspace*{10pt} \vspace*{0.225truein}
\publisher{January 26, 2005}{June 7, 2005}

\vspace*{0.21truein}

\abstracts{
We investigate a surface-mounted electrode geometry for miniature
linear radio frequency Paul ion traps.  The electrodes reside in a
single plane on a substrate, and the pseudopotential minimum of the
trap is located above the substrate at a distance on the order of
the electrodes' lateral extent or separation.  This architecture
provides the possibility to apply standard microfabrication
principles to the construction of multiplexed ion traps, which may
be of particular importance in light of recent proposals for
large-scale quantum computation based on individual trapped ions.
}{}{}

\vspace*{10pt}

\keywords{ion traps, quantum computation, quantum information,
microfabrication, trapped ions} \vspace*{3pt} \communicate{I.~Cirac
\& R.~Blatt}

\vspace*{1pt}\textlineskip    
\section{Introduction}

One promising architecture for quantum information processing (QIP)
is the trapping and manipulation of individual atomic ions in radio
frequency (RF) linear Paul
traps~\cite{ciraczoller,nistpaper,multiplexed,qcionsnist,madsen,roos3qubit,home}.
To become a viable implementation for the execution of large-scale
quantum algorithms, this scheme requires the precise control of many
ions as quantum bits (qubits).  In one possible
implementation~\cite{nistpaper,multiplexed}, ion qubits are held in
separate trapping regions and moved around at will, so that selected
ions may be brought into mutual proximity for the application of
quantum gates and separated again.

\begin{figure}[bt]
\centerline{\epsfig{file=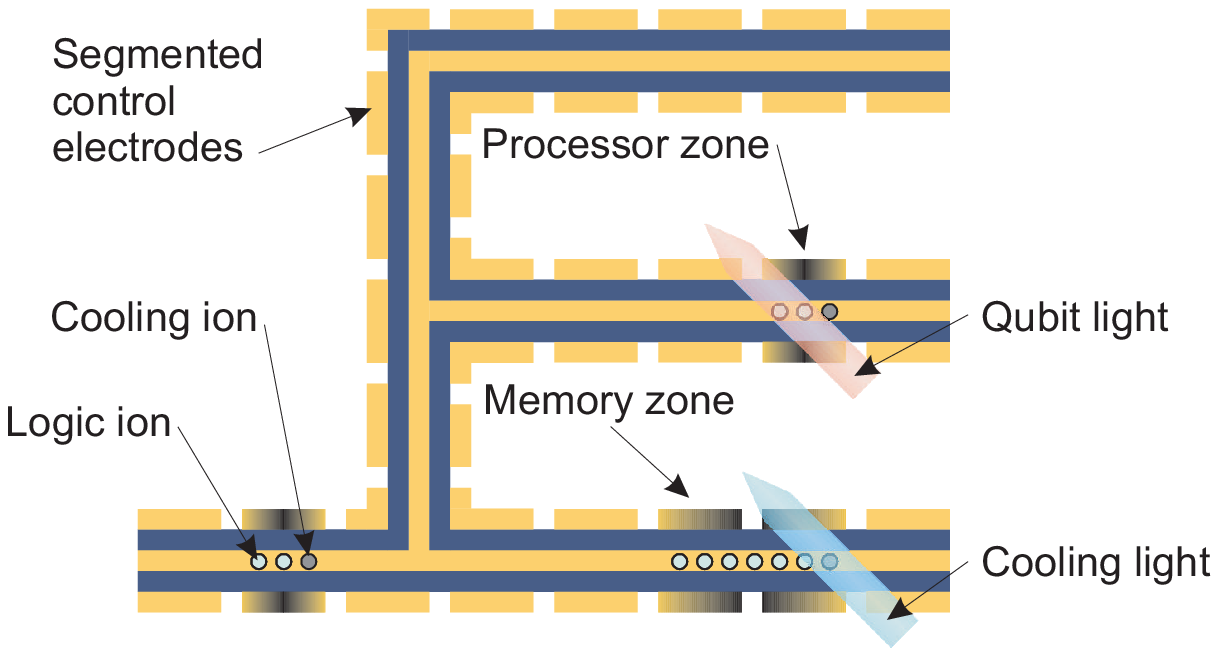, width=0.75 \columnwidth}} 
\vspace*{13pt} \fcaption{\label{largescale} Schematic diagram of
proposed large-scale ion trap array for quantum information
processing~\protect\cite{nistpaper,multiplexed}. Segmented control
electrodes allow logic ions (lighter-colored circles) to be
transferred to and from ``memory'' and ``processor'' regions. After
transport, ions are sympathetically re-cooled by means of cooling
ions (darker-colored circles).  The laser light used for qubit
operations has a frequency different from that used for sympathetic
cooling to minimize decoherence of the qubit ions by the cooling
light. The basic electrode layout depicted here is a ``five-wire''
surface geometry (described below in
Sec.~\protect\ref{sec_surface}); dark-colored electrodes are
radio-frequency (RF) electrodes, and light-colored control
electrodes are held at RF ground.}
\end{figure}

The proposed method~\cite{nistpaper,multiplexed} involves holding
ion qubits in ``memory'' zones of a trap array (see
Fig.~\ref{largescale}).  Ions to be used in a particular gate
operation would be moved from the memory zones to ``processor''
zones.  Quantum gates could be performed on a few ions in such a
processor zone by means of entanglement of the motional modes of
these ions with their internal qubit states as, for example,
in~\cite{ciraczoller,roos3qubit,didi}. Single bit operations can be
performed on ions that have been moved separately into such a
trapping region. For both single- and multi-qubit operations,
individual addressing of the ions through tight laser-beam focussing
would not be required~\cite{multiplexed}. With such a design, highly
parallel processing is possible (there can be many zones with
interchangeable roles), and separate (shielded) regions of the array
may be used for readout of ancilla qubits during a computation so
that the effect of scattered light from this process on memory ions
will be minimized. It appears that rapid ion transportation through,
and ion separation in, such an array leads to heating of the
motional modes~\cite{mary}; therefore the ions will have to be
sympathetically cooled by ``refrigerator'' ions after
transport~\cite{nistpaper,multiplexed,rohde01,blinov02,murray} and
before additional entangling operations. These requirements imply a
physical architecture in which there are many distinct but
interconnected trapping regions. The separation, transportation, and
recombination of ions must be controlled by means of time-varying
potentials on control electrodes linking these trapping regions.

Current linear ion trap
designs~\cite{mary,schrama93,nagerl98,devoe98,barton00,guthohrlein01,hornekaer01,berkeland02}
are somewhat bulky, require considerable time and effort for
construction, and are not easily extendable to more complicated
layouts.  Here we describe an alternative approach, one that employs
methods of microfabrication to construct traps of planar geometry
that are scalable (in both electrode size and trap-array extent) and
somewhat simpler in design.  In Sec.~\ref{sec_bg} we describe the
properties of a typically-used four-rod trap geometry. In
Sec.~\ref{sec_lc}, layered-electrode geometries are discussed, and
in Sec.~\ref{sec_surface} we posit a surface-mounted electrode
architecture to address some of the inherent problems in the other
structures.  Section~\ref{sec_strength} contains a discussion of
trapping parameters for the surface electrode design, and in
Sec.~\ref{sec_imp} we describe a possible implementation using
microfabrication techniques.  We propose a simple benefit of this
design for studies of ion heating due to electrode surface quality
in Sec.~\ref{sec_heat}.

\section{Background}
\label{sec_bg}

\begin{figure}[tb]
\hbox{a) \hspace{0.46 \columnwidth} b)}
\centerline{\epsfig{file=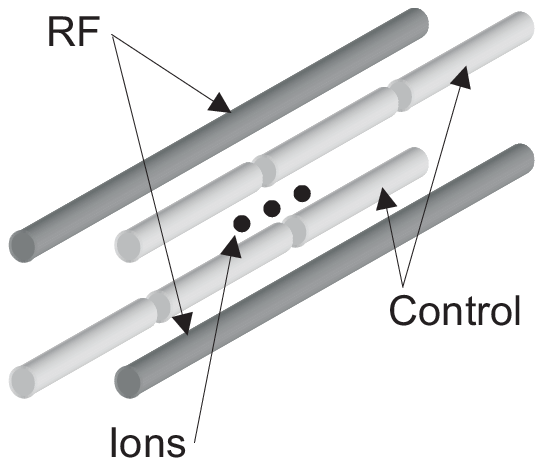, width=0.46 \columnwidth}
\hspace{0.04 \columnwidth} \epsfig{file=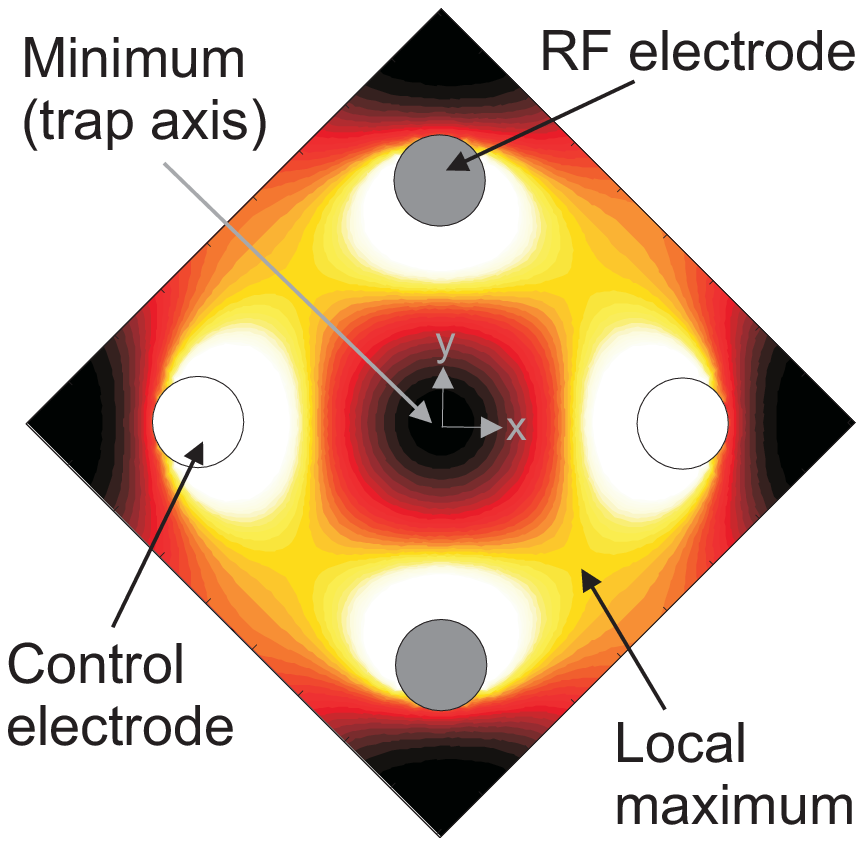, width=0.46
\columnwidth}} \hbox{\hspace{0.2 \columnwidth}  c)}
\centerline{\epsfig{file=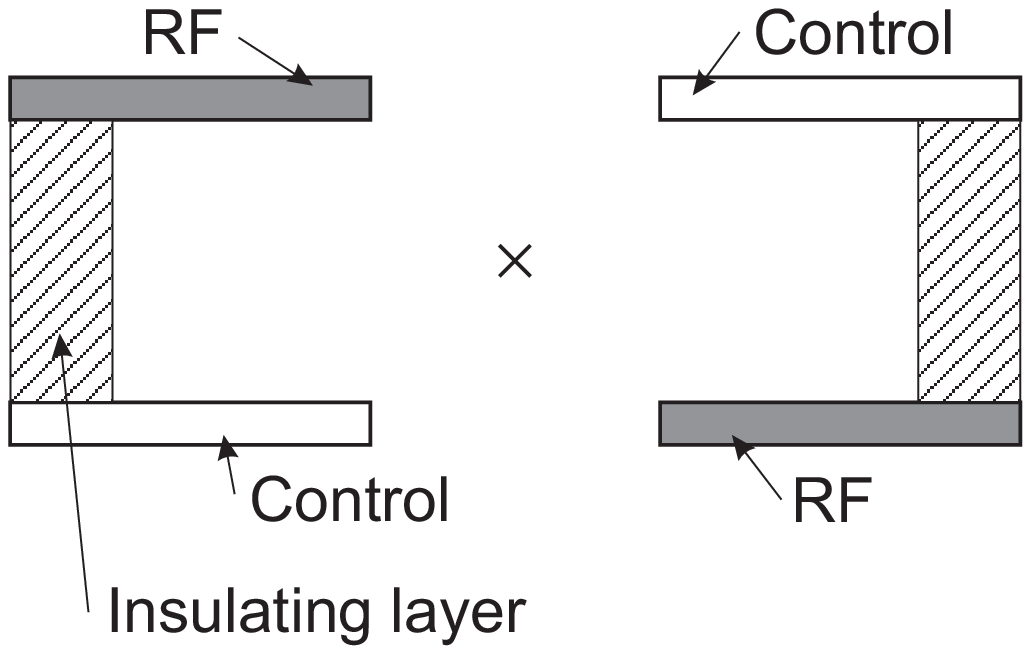, width=0.5 \columnwidth}}
\vspace*{13pt} \fcaption{\label{quad} Standard four-rod, linear RF
quadrupole trap. (a) Perspective drawing of the geometry of this
type of trap. (b) Radial pseudopotential contours (looking down the
axis) of the four-rod trap. Colors from black to white indicate low
to high pseudopotential, respectively. For clarity, contours are not
shown for pseudopotential values above an arbitrary maximum. (c) One
implementation of the four-rod design using two conducting layers
separated by an insulating layer~\protect\cite{mary}.}
\end{figure}

A typical design for a linear RF Paul trap~\cite{paul_rev} consists
of four parallel linear electrodes, each equidistant from the trap
axis and from its two nearest neighbors.  Two of these electrodes
(in opposite positions about the trap axis) are driven by an RF
potential, and the other two ``control'' electrodes are maintained
at RF ground (see Fig.~\ref{quad}). This geometry creates the
required quadrupolar RF field for ponderomotive confinement in the
lateral directions. The two control electrodes are segmented along
their length, and static potentials can be applied to these segments
to provide longitudinal confinement.

The classical motion of a charged particle in such a trap can be
determined by solving the Mathieu equations associated with the
equations of motion of the particle in the applied
potential~\cite{fischer,wuerker}. The parameters of the ion traps
typically employed allow a more intuitive solution. This is the
pseudopotential approximation~\cite{pseud1,pseud2}, in which ions
are bound through a ponderomotive potential

\begin{equation}
U_{2D}({\bf r}) = {Q^2 \over 2 m \Omega^2} \langle E^2({\bf r})
\rangle \label{pseud_eq}
\end{equation}

\noindent at position~${\bf r}$, where $Q$~is the particle's charge,
$m$~is its mass, $\Omega$~is the RF drive frequency (rad/s), and $E$
is the magnitude of the applied electric field. For the standard
quadrupole geometry close to the trap center, this becomes

\begin{equation}
U_{2D}({\bf r}) = {Q^2 V^{2}_0 \over 4 m \Omega^2 R^4} \left( x^2 +
y^2 \right) \label{pseud_quad}
\end{equation}

\noindent if ${\bf r}\equiv x{\bf \hat{x}}+y{\bf \hat{y}}$ is the
displacement from the trap axis. Here~$V_0$ is the peak amplitude of
the RF potential and $R$ is approximately equal to the distance from
the center of the trap to the nearest electrode surface.  This
pseudopotential confines ions in two dimensions, and its curvature
(second spatial derivative) at the trapping location dictates their
oscillation frequency.

For typical trap parameters, one obtains the following for the
motion along either of the two radial directions ($i=\{1,2\}$), up
to a relative phase:

\begin{equation}
x_i(t) \approx X_{i0} \cos \omega_i t \left[ 1 + {q_i \over 2} \sin
\Omega t \right]. \label{motion}
\end{equation}

\noindent In this equation the $x_i$ are the usual coordinates~$x$
and~$y$, the $X_{i0}$ are the pseudopotential oscillation amplitudes
in each direction, and the $\omega_i$ are the pseudopotential
oscillation frequencies (less than half the RF drive frequency). The
parameter $q_i$ is a normalized trap strength

\begin{equation}
q_1=-q_2={2 Q V_0 \over m \Omega^2 R^2} \label{q_eq},
\end{equation}

\noindent with a typical magnitude small compared to~1. Hence the
motion in the radial directions is composed of motion $X_{i0}\cos
\omega_i t$ at frequency $\omega_i$, dubbed the ``secular'' motion,
with excursions of smaller amplitude at the RF frequency. This
superimposed motion is referred to as ``micromotion,'' and its
amplitude is proportional to the ion's radial secular motion
displacement from the trap center.  The pseudopotential picture
effectively ignores ion micromotion, which is added as a
perturbation in Eq.~\ref{motion}.

Figure~\ref{quad} displays equal-pseudopotential contours for a
linear RF Paul trap of standard, four-rod geometry.  The potential
at the trapping region in the center is approximately quadratic with
a constant curvature, independent of direction in the $x$-$y$ plane.
Potential local maxima (saddle points) exist at large displacements
from the center at four locations between the trap electrodes and
may provide escape pathways for energetic ions.  Ions can be readily
trapped and cooled in current traps if these barriers ensure a trap
depth of $\sim\nobreak1$~eV or larger.

Current and past implementations of this four-rod design have
involved, for example, macroscopic rods in a quadrupole
formation~\cite{rodtrap,raizen} and a quadrupole formed by the
inside edges of slots cut into pairs of alumina wafers~\cite{mary}
as well as other
designs~\cite{schrama93,nagerl98,barton00,guthohrlein01,hornekaer01,berkeland02}.
All of these devices involve macroscopic machining of some sort.
Such traps function well, but they do not appear to offer a path
toward the smaller, more complicated devices that will be necessary
for large-scale QIP with trapped ions.

\section{Layered Construction}
\label{sec_lc}

\begin{figure}[tb]
\hbox{a) \hspace{0.46 \columnwidth} b)}
\centerline{\epsfig{file=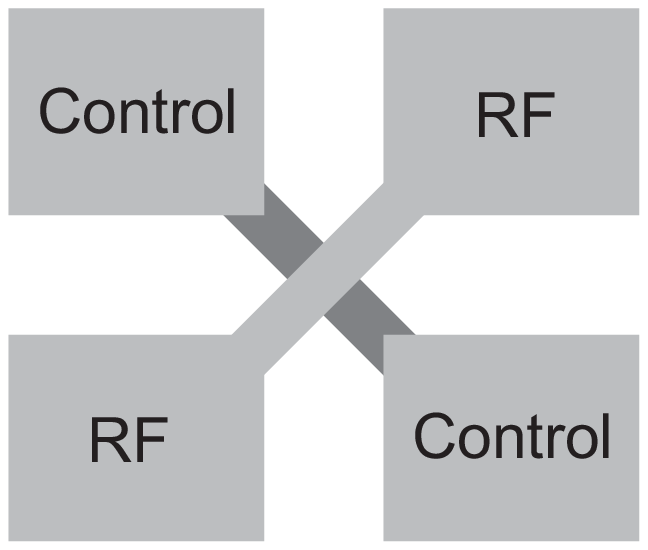, width=0.46 \columnwidth}
\hspace{0.04 \columnwidth} \epsfig{file=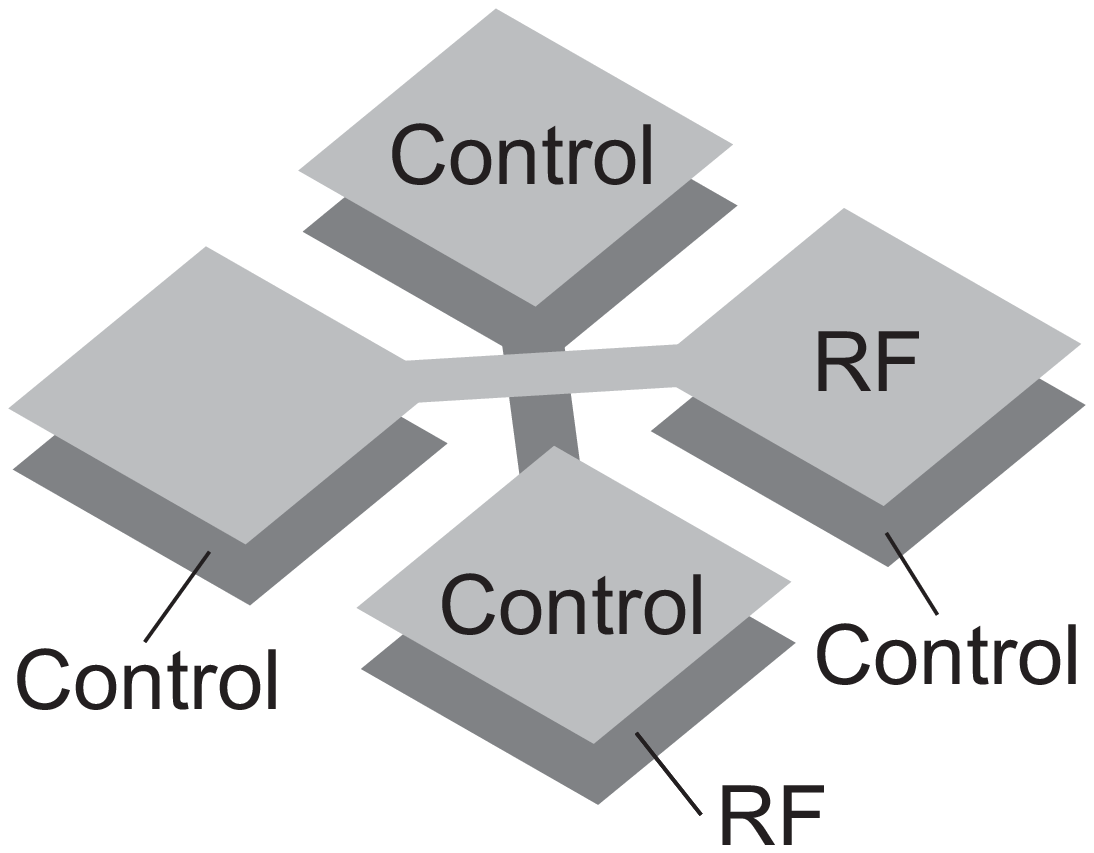, width=0.46
\columnwidth}} \vspace*{13pt} \fcaption{\label{bridged_cross}
Schematic diagram of modified two-layer cross to maintain a trapping
potential at the center of the cross.  Away from the cross, each of
the four linear trapping regions consists of a four-rod linear trap
implemented with two conducting layers as in Fig.~\ref{quad}c. The
ions would be trapped halfway between the two layers. (a) Top view
showing diagonal bridges that connect opposite RF electrodes in both
layers. Bottom layer is dark gray. Alternatively, opposing control
electrodes can be connected. (b) Perspective view.}
\end{figure}

A straightforward approach to incorporating microfabrication into
trap construction is to use multi-layer processing to reproduce the
quadrupole geometry on a smaller scale. Although the standard
four-electrode design may be implemented with two electrode layers
and an intervening insulating layer as in~\cite{mary} (see
Fig.~\ref{quad}c), this construction is not adequate for tee
junctions and other complicated patterns. However, four-junction
crosses can be made with a slight modification to the two-layer
design. Figure~\ref{bridged_cross} shows the proposed modification
to a standard two-layer cross (the intersection of two linear traps,
each of which is composed of two conducting layers) such that
charged particles can be trapped in the intersection region. By
connecting the RF electrodes (or alternatively the control
electrodes) with diagonal bridges, a trapping minimum can be created
at the cross's center. Variation of the potentials on control
electrode segments (not shown) allows movement of charged particles
through the cross region into any of the branches while maintaining
a stable minimum at every point along the trajectory.

\begin{figure}[tbp]
\hbox{a) \hspace{0.46 \columnwidth} b)}
\centerline{\epsfig{file=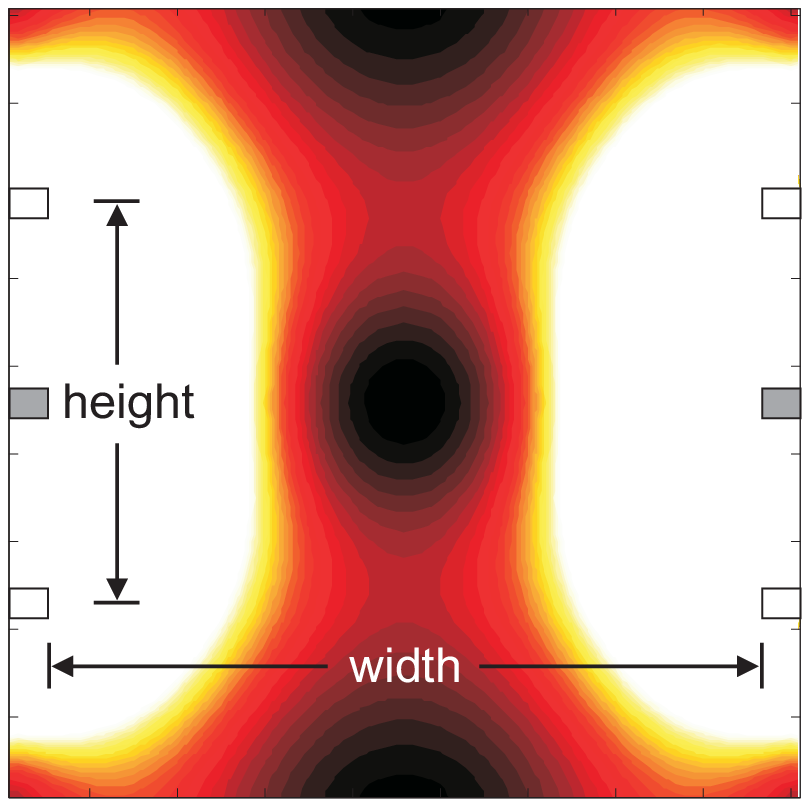, width=0.44
\columnwidth} \hspace{0.04 \columnwidth}
\epsfig{file=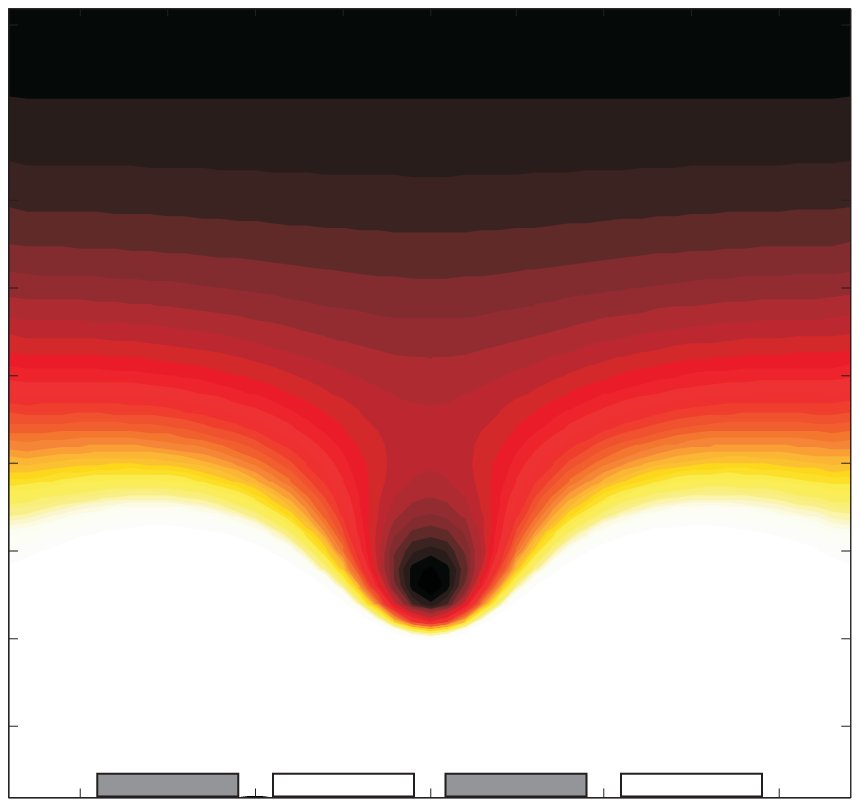, width=0.44 \columnwidth}} \hbox{c)
\hspace{0.46 \columnwidth}  d)}
\centerline{\epsfig{file=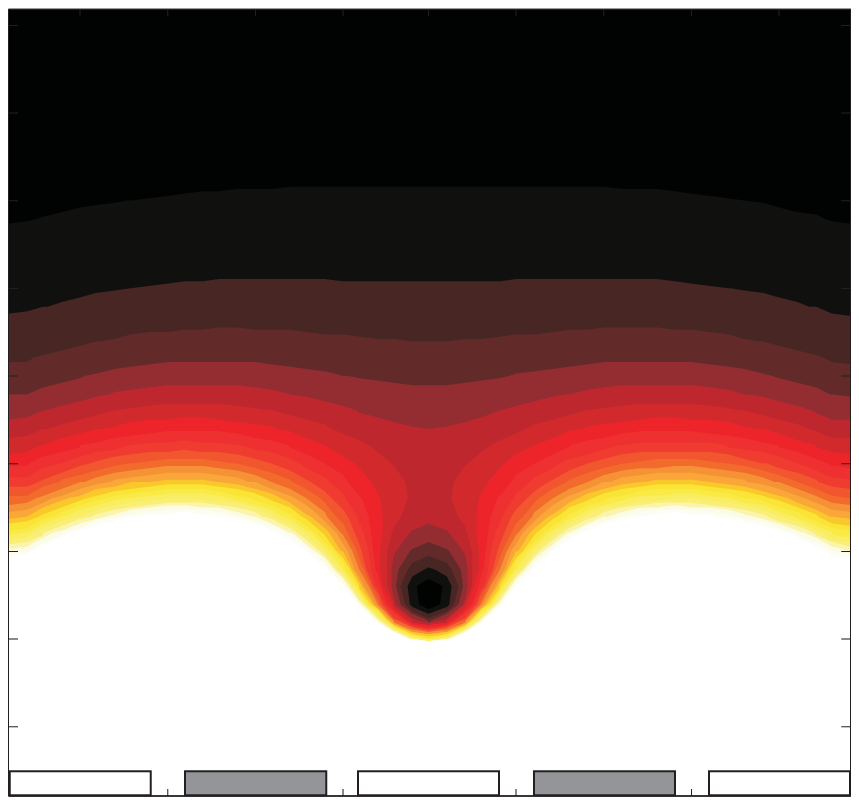,width=0.44 \columnwidth}
\hspace{.04 \columnwidth} \epsfig{file=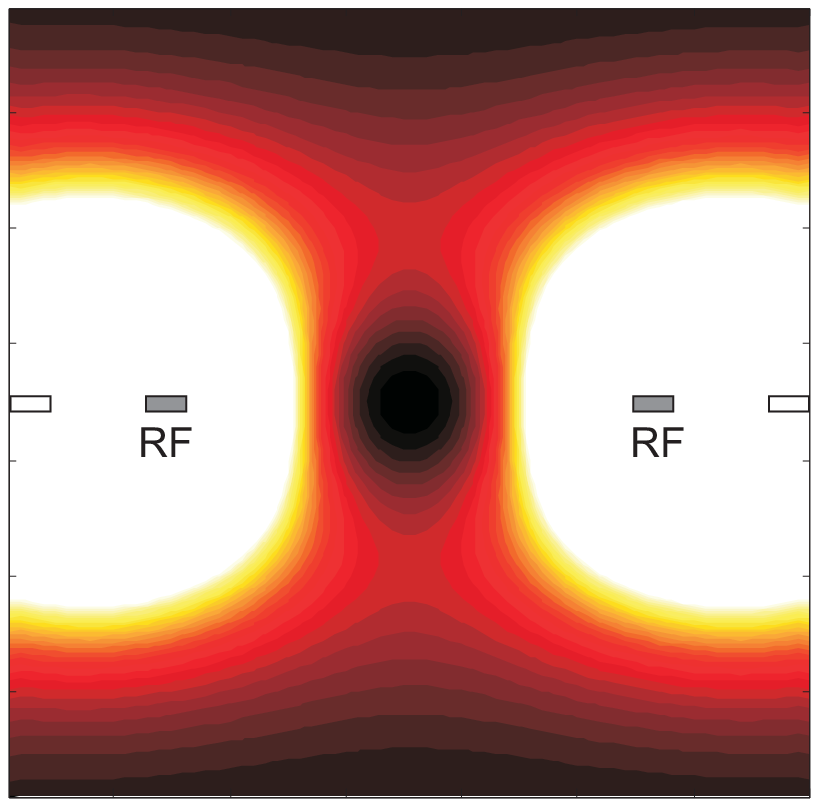, width=0.44
\columnwidth} } \vspace*{13pt} \fcaption{\label{trap_pseud} Radial
pseudopotential contours (looking down the trap axis) for four
linear RF-trap geometries. Colors from black to white indicate low
to high pseudopotential, respectively.  For clarity, contours are
not shown for pseudopotential values above an arbitrary maximum. An
RF potential is applied to the gray electrodes, and the white
electrodes are held at RF ground. For engineering simplicity, it
would usually be more convenient to apply the control potentials to
the RF-ground electrodes.  For each geometry, secular
frequencies~$f_i$ and trap depths~$u_{\rm max}$, normalized to the
values for a two-layer quadrupole with an aspect ratio (equal to the
width over the height)~$\gamma=1$ (Fig.~\protect\ref{quad}c) with
identical drive frequency, RF potential amplitude, and ion mass, are
listed. These values are for identical distance~$d$ from the trap
axis to the nearest electrode surface and for an electrode thickness
of~$0.10d$, and they were calculated numerically as described in the
text. (a) A three-layer geometry with~$\gamma=1.8$.  For this
geometry $f_i=0.52$ and $u_{\rm max}=0.078$. For the simulation of
this design, the electrodes were taken to extend semi-infinitely in
the lateral directions. (b) A planar geometry with trapping axis
offset from the plane of the electrodes. For this geometry
$f_i=0.34$ and $u_{\rm max}=0.017$. (c) A similar trapping potential
produced with a surface geometry made up of five electrodes. For
this geometry $f_i=0.30$ and $u_{\rm max}=0.010$. (d) A planar
geometry with trapping minimum in the plane of the electrodes. For
this geometry $f_i=0.32$ and $u_{\rm max}=0.051$.}
\end{figure}

As an alternative to the two-layer realization of the four-rod
design, a deformation of the quadrupole to six electrodes in a
three-layer geometry allows more flexibility with trap layout (see
Fig.~\ref{trap_pseud}a)~\cite{multiplexed,deslauriers}. This
construction requires three conducting layers sandwiching two
insulating layers.   A potential problem with this approach may be
the difficulty of fabrication using standard microfabrication
techniques.  The insulating layers must be low RF-loss materials,
and ideally, they must have a thickness approximately equal to the
lateral electrode spacing.  This dimension must be $\gtsim
10$~$\mu$m to allow for optical access. For aspect
ratios~$\gamma\equiv$\ width/height (Fig.~\ref{trap_pseud})
between~1 and~15, the depth of the trap, due to the nearby local
maxima above and below the trapping region, scales as approximately
$\gamma^{-2.01}$ for a constant trap width (this assumes conductors
rectangular in cross-section with thickness approximately~0.02 times
the trap width). Similarly, the secular frequency obtained for a
constant applied potential at a fixed frequency with the same
electrode geometry scales as approximately $\gamma^{-0.93}$ in this
range of aspect ratios.

These scaling relations were determined numerically by means of
pseudopotential calculations. This and the other simulations
referred to below relied on numerical relaxation to determine the
electrostatic potential of an electrode geometry. The field was
determined from the potential, and the pseudopotential was
calculated as in Eq.~\ref{pseud_eq}.

The scaling relations determined for the three-layer constant-width
geometry suggest that the insulating layers should be quite thick,
and to facilitate fabrication, methods are required for either
deposition or deep etching of thick layers. This limits the
materials and processes that can be used. Typical materials such as
silicon and gallium arsenide lead to significant RF loss; aluminum
gallium arsenide, which can be grown epitaxially on gallium
arsenide, has been suggested as an insulator~\cite{madsen}, however.
Thick layers of low-loss materials such as sapphire, alumina, and
quartz are difficult to deposit and etch. Lateral extension of the
center electrodes in the three-layer geometry beyond the upper and
lower electrodes towards the center of the trap~\cite{schrama93}
strengthens and deepens the trap, but fabrication of such a
structure may require more etch steps and hence may become more
difficult.

The problems with designs of high aspect ratio ($\gamma \gg 1$) are
not as pronounced for the standard four-electrode
configuration~\cite{qcionsnist,madsen,mary}, and small versions of
this type of trap may be built using multi-layer microfabrication.
Two conducting layers and one insulating layer, each as thin as a
few micrometers, would be suitable for a trap.

\section{Surface Geometry}

\label{sec_surface}

Deformation of the standard quadrupole geometry electrodes into a
plane preserves a quadrupolar field for a wide range of parameters
while also providing an avenue to more straightforward fabrication.
The most obvious planar deformation is to move one control electrode
and one RF electrode into the plane of the remaining two electrodes,
such that the electrodes are of alternating potential (RF, control,
RF, and control), as shown in Fig.~\ref{trap_pseud}b. The trap axis
remains in approximately the same position, above and between the
center two electrodes (there is another trap axis symmetrically
opposite this one, beneath the electrodes, but this axis will
typically reside within the substrate).  This ``four-wire'' geometry
appears promising, but making connections to the segments of both
control electrodes requires multi-layer processing or the use of
vias through the substrate.  Alternatively the outer RF and control
electrodes could be segmented and different control potentials
applied, but this is complicated by the practical difficulty of
applying control potentials to the electrodes without affecting
application of the RF potential. An alternative is the segmentation
of only the outer control electrode for axial confinement. The
asymmetry in this configuration requires adjustment of the other
(nonsegmented) control electrode to compensate for the lateral field
due to the single segmented electrode. In addition, structures such
as tee junctions may be difficult to construct with four wires in
this configuration.

To alleviate these problems with the four-wire planar geometry, the
four-rod quadrupole can instead be deformed into a ``five-wire''
design, in which the center and outer electrodes are maintained at
RF ground, and the remaining two electrodes support an RF potential
(see Fig.~\ref{trap_pseud}c). In this geometry, the trap axis is
above (and below) the center electrode, and the two outer electrodes
can be segmented for longitudinal confinement and control.  There
may be advantages for ion separation if the center electrode is
segmented (see below), but making connections to the segments would
require multi-layer processing or through-layer vias, a complication
we do not consider here.  As the five-wire design obeys a mirror
symmetry through the plane containing the center electrode and
perpendicular to the substrate plane, tee junctions are
straightforward to construct (see Fig.~\ref{largescale}).

\begin{figure}[tb]
\hbox{\hspace{.02 \columnwidth} a) \hspace{.28 \columnwidth} b)
\hspace{.28 \columnwidth} c)} \centerline{
\epsfig{file=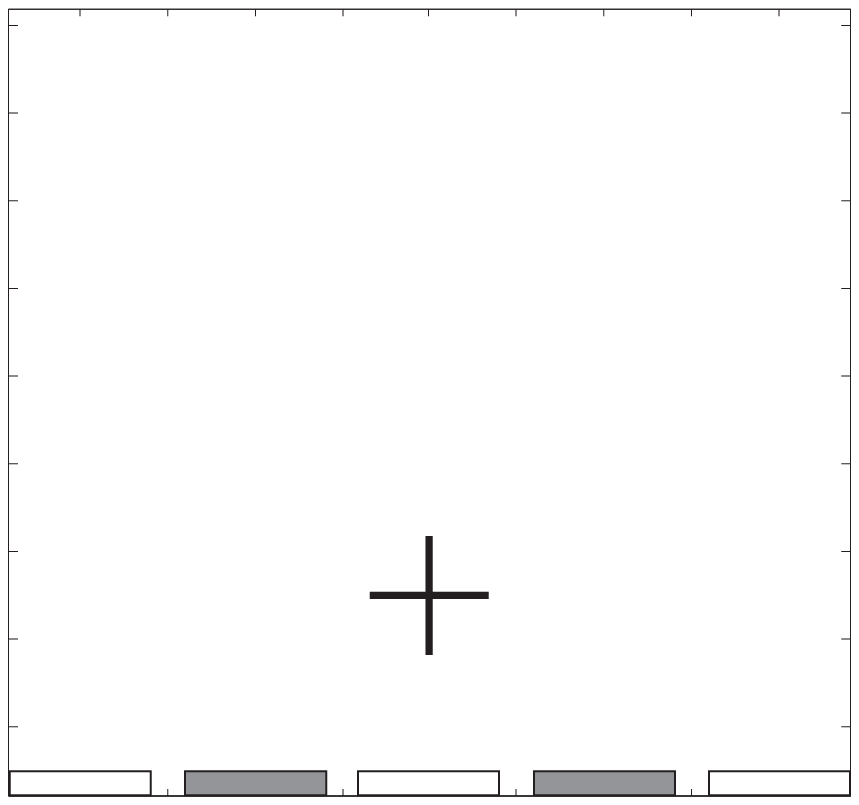, width=0.27 \columnwidth}
\hspace{0.03 \columnwidth} \epsfig{file=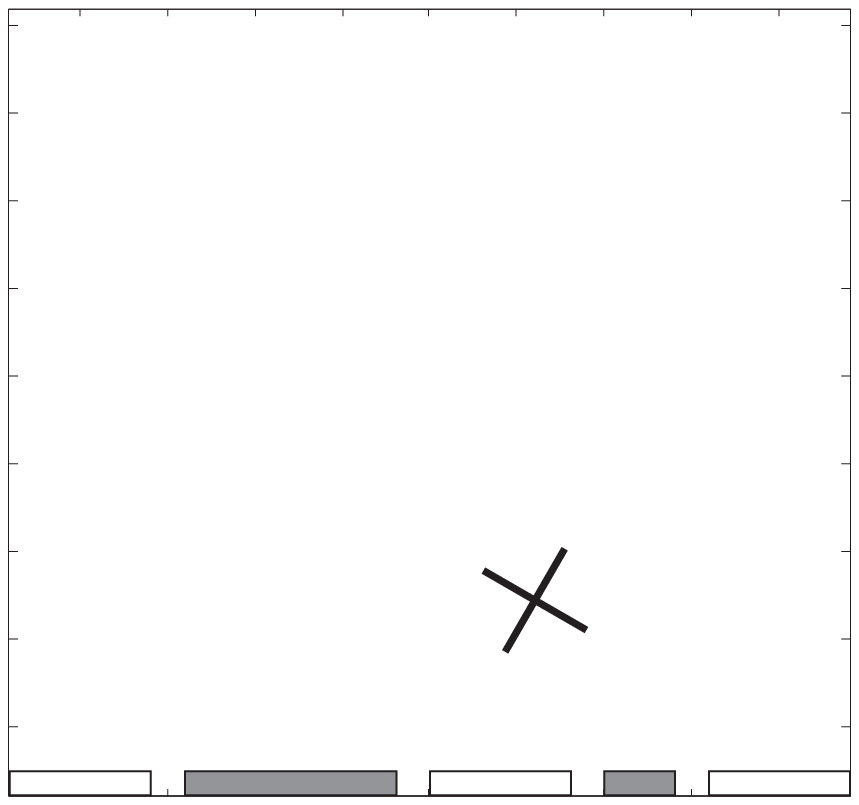,
width=0.27 \columnwidth} \hspace{0.03 \columnwidth}
\epsfig{file=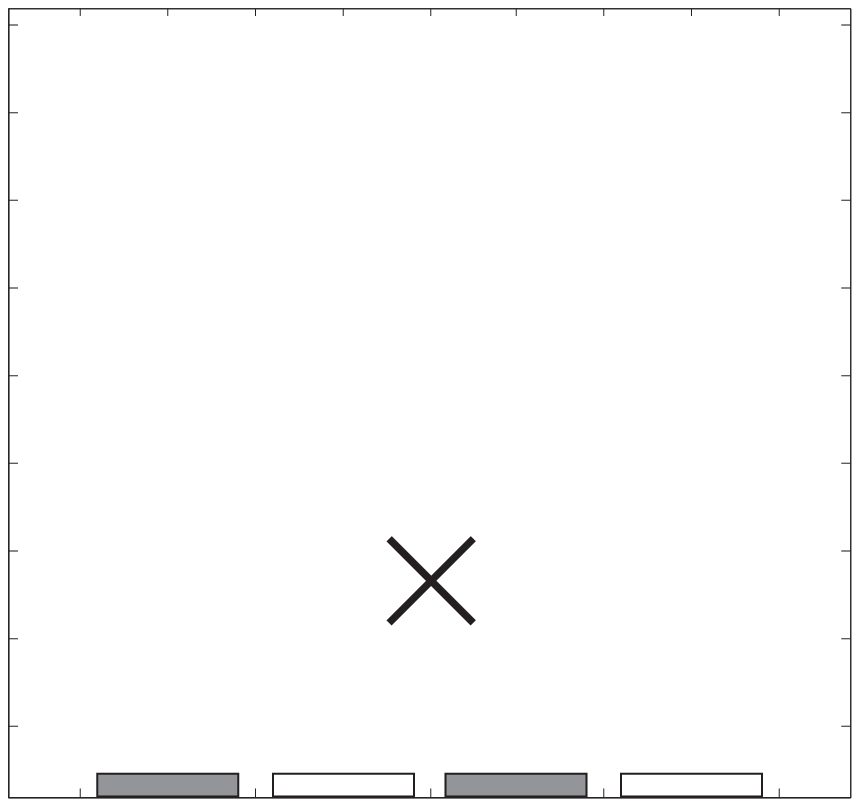, width=0.27 \columnwidth}}
\vspace*{13pt} \fcaption{\label{trap_axes} Trap principal axes
directions for three different surface trap configurations.  The
third principal axis is perpendicular to the page and not shown. The
transverse axes directions are depicted by two orthogonal lines at
the trap axis location and are determined by static potentials
applied between segments of the gray (RF) and white (control)
electrodes. The axes directions are close to those shown for a large
range of control voltages (relative to the RF potential amplitude).
(a)~Five-wire geometry, all electrodes of equal width. (b)~Five-wire
geometry, RF electrode widths altered. (c)~Four-wire geometry.}
\end{figure}

In a linear trap, an added static potential, as opposed to the
pseudopotential, defines the two radial major trap axes, because the
pseudopotential is cylindrically symmetric near the center of the
trapping region (see Eq.~\ref{pseud_quad} and Fig.~\ref{quad}b).  In
the designs discussed here, the static potential, acting alone,
leads to a maximally trapping direction and a perpendicular,
maximally anti-trapping direction in the $x$-$y$ plane
(perpendicular to the long direction); these two directions define
the major trap axes for the combination of the static potential and
the RF pseudopotential. By symmetry, the standard five-wire design
(Fig.~\ref{trap_pseud}c) provides a static potential with major trap
axes perpendicular and parallel to the substrate (the third axis in
a linear trap will be along the long dimension of the wires) as in
Fig.~\ref{trap_axes}a. By changing the relative lateral sizes of the
electrodes, the major axes of the trap may be rotated (for instance,
if the width of one of the RF electrodes is increased by half its
width, and the width of the other RF electrode is decreased by the
same amount, the principal axes rotate by~30$^\circ$) as depicted in
Fig.~\ref{trap_axes}b. This allows Doppler-cooling laser beams whose
wavevectors are parallel to the substrate to cool the ions' motion
in both radial-axis directions~\cite{itano82}, that is, the ions
will not be heated through recoil along any of the principal axes.
This might be necessary as beams with wavevectors nonparallel to the
surface may produce significant amounts of scattered light. The
orientation of the trap principal axes in the four-wire surface
geometry is approximately~$\pm45^{\circ}$ from horizontal (parallel
to the substrate) for a large range of electrode potentials and thus
allows laser-cooling along all three axes with a beam parallel to
the surface (see Fig.~\ref{trap_axes}c).

A potential benefit of the planar electrode configuration is that
the location and strength of the trapping region may be altered by
varying the electrode structure in one dimension (the lateral
direction) only. Through overall scaling of the electrode size and
spacing, the distance from the trap axis to the electrodes and
substrate may be controlled so that the effects of surface proximity
on ion heating~\cite{turchette00} may be studied (see
Sec.~\ref{sec_heat}). This distance can be changed continuously
along the length of the electrodes, in effect creating a trapping
region for any desired surface-ion separation distance.

\begin{figure}[tb]
\centerline{\epsfig{file=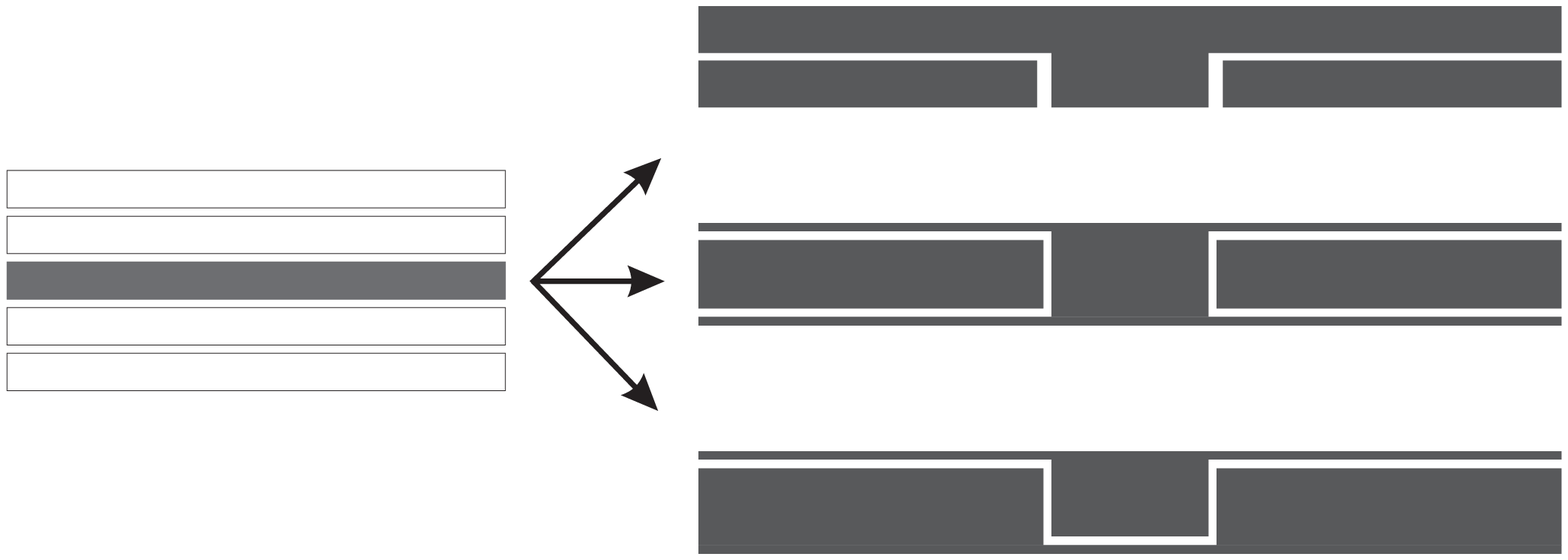, width=0.95 \columnwidth}}
\vspace*{13pt} \fcaption{\label{split} Three different options for
layout of composite center control electrode in planar five-wire
geometry to allow for efficient separation of ions. A similar layout
can be used in the four-wire geometry, as well. Electrode structure
appears as viewed from above the substrate. The top two versions
split the center electrode into three electrodes: the effect of the
continuous electrode in each case can be partially canceled by the
shorter electrodes everywhere but in the separation region. In the
bottom version, the center electrode is split longitudinally in two
continuous electrodes, and the separation potential can be applied
to the electrode with greater extent in the separation region. The
field due to this electrode is nearly canceled outside this region
by the other, relatively wider electrode.}
\end{figure}

Another possible benefit of the surface geometry is that it may
facilitate ion separation~\cite{mary,teleportation}.  Two ions or
groups of ions in the same longitudinal trapping region may be
separated by raising the potential of an electrode near the center
of the trapping region. As this potential is raised, the ions'
motional frequencies necessarily drop as the external potential
becomes flat. The minimum motional frequency during the separation
increases as the splitting electrode's size and the ions' proximity
to this electrode become smaller.  According to numerical
simulation, the minimum frequency~$\omega_{\rm min}$ is
approximately proportional to~$1/L$ (where $L$ is the distance from
the ions to the nearest control electrode) in agreement
with~\cite{home}. We want to keep this minimum frequency as high as
possible, both to minimize kinetic excitation of the ions' motional
modes and motional heating due to stray fluctuating charges, and to
maximize the speed at which the operations can be
performed~\cite{home,mary}. In the five-wire geometry, the center
electrode will be nearest to the ions (in the four-wire geometry,
the inner control electrode will be at a comparable distance from
the trap axis for a similarly sized geometry). This central
electrode may be split into several segments, some with greater
extent in the longitudinal or lateral directions (see
Fig.~\ref{split}). Potentials applied to a subset of these
electrodes may be tuned to nearly cancel the field due to the
remaining electrodes everywhere but in a small region where the ions
will be separated. The close proximity of the electrode can lead to
very efficient ion separation.

Another possible four-wire configuration that will create a
quadrupole field is a pair of parallel RF electrodes flanked by two
outer RF-ground electrodes. Here the trap axis will lie between the
two RF electrodes, as shown in Fig.~\ref{trap_pseud}d. This places
the trap very close to the substrate unless a slot is etched through
it between these two electrodes. It also places the RF electrodes
between the trap axis and the control electrodes, limiting the
effectiveness of the control electrodes in the longitudinal
direction. An alternative is the use of freestanding wires for this
design, with separate electrodes for axial confinement.

\section{Strength and Depth of Surface-Electrode Geometry Trap}
\label{sec_strength}

The trap frequencies in both surface trap designs can be made
comparable to those in the standard quadrupole of similar size.  All
else being equal, the radial frequencies are approximately one sixth
to one third (depending on exact electrode shape) of those in a
standard four-rod quadrupole of similar size (see Appendix~A for a
comparison to a thin-wire surface design).  However, the surface
traps suffer from a much shallower trap depth. There is a local
maximum in the pseudopotential at a distance proportional to and of
the same order as~$d$ above the trap axis in both cases (see
Figs.~\ref{trap_pseud}b,~\ref{trap_pseud}c, or
Fig.~\ref{analytic_layout}), and the depth of the trap (defined by
this maximum) is smaller by a factor of approximately~30 to~200 than
the depth of the standard quadrupole trap, depending on electrode
shape. Typical trap depths for the four-rod quadrupole design are on
the order of a few electron volts (1~eV is approximately equivalent
to~$12\,000$~K); a trap depth significantly less than this is
adequate once the ion is trapped and cooled, but the initial loading
efficiency could be compromised due to the small trap depth. Ions
are typically produced by means of electron-impact ionization or
photoionization of a thermal neutral atomic source. Some typically
trapped atomic species, such as beryllium, must be produced from an
evaporation source at a temperature of $\sim\nobreak1000$~K. One
will need to depend on the low-energy tail of the Boltzmann
distribution of atom kinetic energies for trappable atoms if the
trap depth is low. A possible solution is the integration of a
load/reservoir trap of a more standard design. This trap would be
more efficient for loading ions, and a large number can be obtained
and held until they are needed, at which time they can be shuttled
into the surface trap for computational purposes. Also, lost ions
(due to chemical reaction or background-gas collisions) can be
replaced with ions from this reservoir. Numerical simulations
suggest that a smooth crossover between the two trap geometries is
feasible.

The pseudopotential trap depth in the surface designs can be
increased by adjusting the static potential used to bring about
axial confinement. The static potential due to voltages applied to
the segmented electrodes is typically trapping in the direction of
one of the radial principal axes and antitrapping in the direction
of the other.  Calculations suggest that variation of the principal
axes angles (as mentioned above) and the magnitude of the control
voltages can deepen the trap by factors of about~2.

Although current trap designs must be reduced in size to scale to
large numbers of qubits, there may be an ultimate limit to how small
any linear RF Paul trap can be made. Calculations (see Appendix~B)
suggest that scaling beyond an ion-electrode distance of tens of
nanometers may become impossible due to reduced trap depth, even
using a standard quadrupole trap.

\section{Implementation}
\label{sec_imp}

\begin{figure}[tb]
\centerline{\epsfig{file=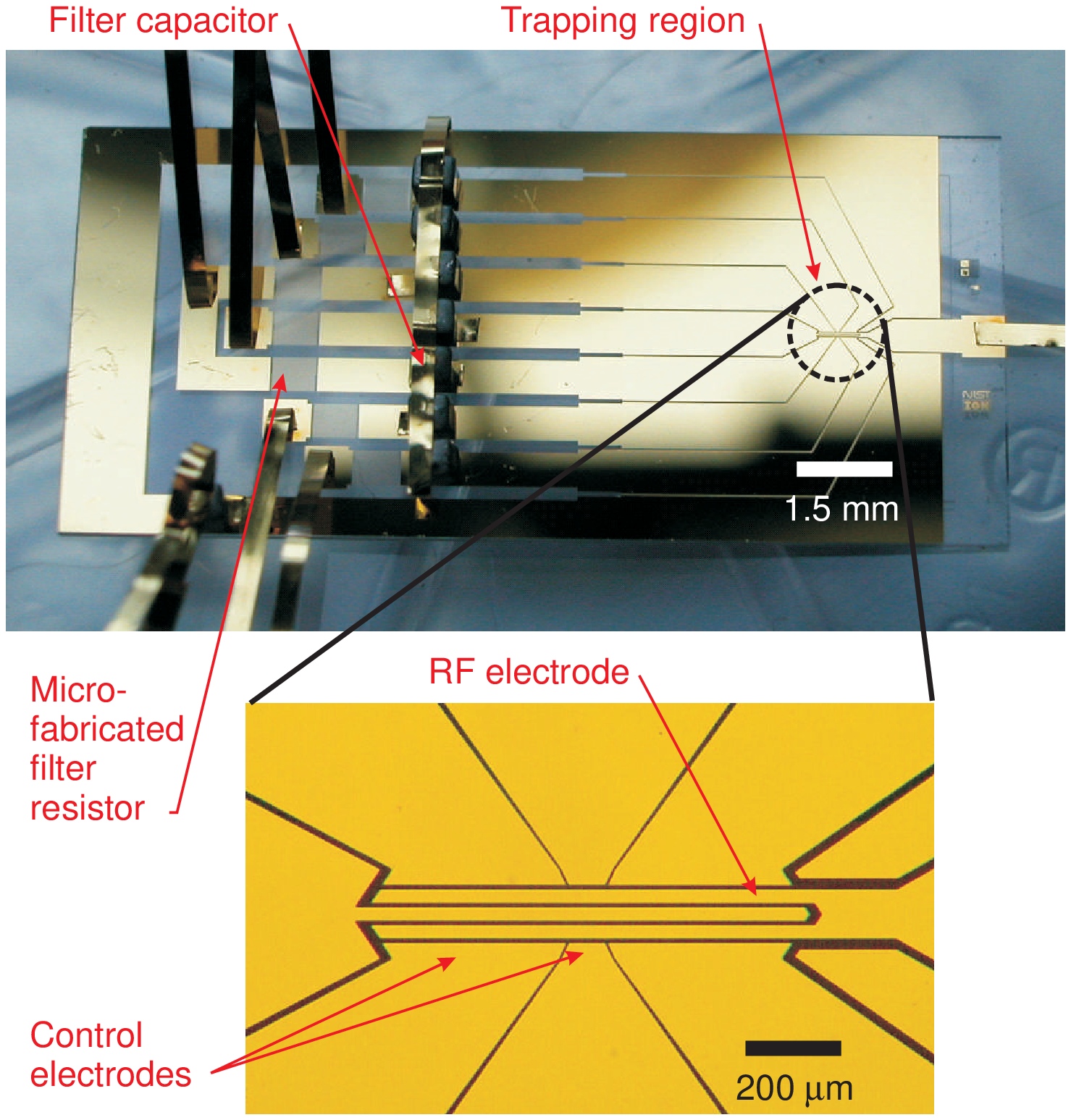, width=0.8 \columnwidth} }
\vspace*{13pt} \fcaption{\label{pics} Micrographs of a five-wire,
one-zone linear trap fabricated of gold on fused silica. The top
figure is an overview of the trap chip showing contact pads, onboard
passive filter elements, leads, and trapping region. Substrate
dimensions are 10~mm~$\times$~22~mm~$\times$~500~$\mu$m. The lower
image is a detail of the trapping region indicated by the dotted
ring in the top figure. The substrate material appears dark-colored
in the lower image.}
\end{figure}

Utilizing the techniques developed for the microelectronics
industry, ion traps of planar geometry can be constructed in only a
few fabrication steps.  The basic structure can be made by means of
patterned metal deposition onto an insulating substrate.
Photolithography or electron beam lithography can be utilized to
define the electrodes. This method is similar to that used for
neutral atom optics and atom guides on
chips~\cite{hinds99,lau99,folman00,prentiss00,reichel01,muller01,lev03b}.
An ion trap structure fabricated in this manner is shown in
Fig.~\ref{pics}.

Calculations suggest that a trap of this construction will be
capable of trapping ions approximately 50~$\mu$m from the chip
surface.  Secular frequencies will be in the 10~MHz range for an
applied RF potential of amplitude 100~V at 100~MHz. Control
electrode potentials of a few volts will lead to axial frequencies
in the few megahertz range.  The dissipation in the trap can be
modeled as a resistor representing the leads in series with the trap
capacitance and its equivalent series resistance, which is
proportional to the dielectric loss.  In this case, the dissipated
power will be $P=I_{\rm RMS}^2 (R_{\rm Lead} + R_{\rm ESR})=(C
\Omega V_{\rm RMS} )^2 (R_{\rm Lead} + \tan\delta/(C \Omega))$,
where $I_{\rm RMS}$ is the root-mean-square RF current, $R_{\rm
Lead}$ is the total lead resistance, $R_{\rm ESR}=\tan\delta/(C
\Omega)$ is the equivalent series resistance of the capacitance $C$
between the RF and RF-grounded electrodes, $V_{\rm
RMS}=V_0/\sqrt{2}$, and $\tan\delta$ is the loss due to the
substrate.  For the trap parameters listed above, lead resistance
totalling 1~$\Omega$, a total trap capacitance of 3~pF, and a loss
tangent of 0.0004 (for fused silica), the dielectric loss is
slightly less than the resistive loss, and the RF power dissipation
in the trap in Fig.~\ref{pics} is 20~mW (see Appendix~C for a list
of material and electrical properties for typical and suggested
substrates).  The values for the lead resistance and capacitance are
correct for the structure shown in Fig.~\ref{pics}, but will vary
with the overall size of the trap and the number of electrodes.

The control electrodes require RF grounding capacitors, and it is
desirable to filter the external control potentials with low-pass
RC~filters. We have typically accomplished this with a separate
board containing surface-mounted components and another set of
connections between this ``filter board'' and the trap~\cite{mary}.
Typical values for the resistor and capacitor are approximately
1~k$\Omega$ and 1~nF, respectively, leading to a high frequency
roll-off of approximately 160~kHz. With these components, control
electrode voltages can be changed such that ions can be transported
in segmented traps in tens of microseconds. Future traps may need to
employ filters with higher roll-off frequencies so that ions can be
transported more quickly.

Since the fabrication procedure for the planar trap is relatively
straightforward, passive filter elements may be incorporated
directly on the trap chip.  The chip shown in Fig.~\ref{pics}
contains a low-pass filter on each control electrode lead consisting
of a resistor and capacitor to ground. The resistor is a long, thin
meander wire of the same metal as the trap electrodes, and the
capacitor is a separate surface-mount component. Capacitors could be
made during the microfabrication, as a pair of parallel plates with
silicon dioxide as an insulator. This construction adds two
thin-film depositions to the fabrication.

One concern with this design is the proximity of insulating surfaces
between the electrodes that can hold charge and perturb the trapping
field in the vicinity of the ions. In the proposed surface electrode
configuration, the spaces between the electrodes will be bare
insulator, and charges on these surfaces may be problematic.  To
minimize the effect of charging, trenches that undercut the metal
layer may be etched into the substrate between the electrodes; if
the depth of the trenches is more than a few times the width of the
space between the electrodes, the (possibly charged) insulator's
influence on the field at the ion will be suppressed. Toward this
end, the insulator may be etched via a wet chemical etch or a
reactive ion etch after an additional lithographic masking step. An
alternative is to electroplate the electrodes to a height more than
a few times the inter-electrode spacing to accomplish the same
effect.

The choice of materials used in the construction of these traps is
important as the electrode surface can influence vibrational-mode
heating of the ions (see next section), and the substrate supporting
these electrodes must have low RF loss.  The planar trap
fabrication, as described above, may incorporate many different
metals, as evaporation, sputtering, and electroplating can employ a
wide variety of materials.  Gold's relative surface cleanliness
makes it a reasonable choice, although refractory metals or other
noble metals that do not support significant surface oxides may also
be investigated. Similarly, there are many choices for substrate
material, as the fabrication method requires only a relatively
smooth surface and reasonable adhesion properties to metal films.
Fused silica and quartz have the added benefits of low RF loss and
some ability to be etched. Other options include aluminum nitride,
sapphire, alumina, and diamond, although etching difficulties with
the latter three materials may imply use only in the electroplating
procedure. Doped silicon may also be used as an electrode material,
and the electrodes can be defined using through-wafer reactive ion
etching (ions have been trapped in a doped silicon structure of the
standard quadrupole design~\cite{britton}). There would be no
insulator near the trapped ion's position, and optical access may be
provided through etched gaps between the electrodes in the planar
structure. The doped silicon electrodes can be supported by a glass
substrate (away from the trapping region) attached by means of
anodic wafer bonding~\cite{silicon_trap}. Gallium-arsenide and
aluminum-gallium-arsenide combinations could also be good electrode
choices~\cite{chrisgaas}, and in a separate design, silicon carbide
has been suggested to provide structural support for metallic
electrodes~\cite{benkish}.

The effect of a dielectric substrate on the trapping fields above
the surface-mounted electrodes must be considered, as many of the
suggested materials have relative dielectric constant $\epsilon
> 10$. The field in the dielectric will be modified, and at the
dielectric-vacuum interface, a component of the electric field will
be discontinuous. This may lead to modification of the field at the
location of the ions.  Simulations of the surface geometry with a
substrate with $\epsilon = 1$ (which is the case for all the field
simulations discussed above) have been compared to similar
simulations with a substrate with $\epsilon = 10$.  For gaps between
the electrodes as big as 1/4 the width of the electrode segments,
the difference in the electric field in and near the trapping region
is on the 1~\% level.  This difference will be smaller for the
smaller relatively-sized gaps employed in current designs, but it
could be significant in the limit of thin wire-electrodes separated
by large open regions of the substrate surface.

We also note that the proximity of the electrodes to the ions
possible with the surface-electrode design suggests the possibility
of performing single-qubit rotations by use of microwaves delivered
to the ions via the trap structure.  A subset of the planar trap
wires could make up a waveguide or cavity for long-wavelength
radiation. Unlike coherent operations performed via two-photon
stimulated-Raman transitions, single-bit rotations brought about
using microwaves are practically immune to the effects of
spontaneous emission~\cite{nistpaper,wund}.  Such manipulation of
electronic hyperfine levels using near-field microwave radiation has
been proposed for ion traps~\cite{wund} and neutral-atom
chip-traps~\cite{treutlein}.

\section{Heating Studies}

\label{sec_heat}

Heating of a trapped ion's motional modes can occur from thermal
electronic noise in the metal of the electrodes (or in resistive
elements connected to these electrodes), fluctuating patch
potentials on the surfaces of the electrodes, and perhaps other
unknown sources. Each creates heating rates that scale as some power
of the distance between the ion and the
electrodes~\cite{turchette00,henkel99}. The planar trap described
above offers a method to study the effects of this type of heating
and to verify the scaling with distance.

Heating due to Johnson noise in the electrodes will cause
fluctuations in the field at the ion, and the heating rate due to
these fluctuations is expected to scale as $d^{-3}$ for $d \ll
\delta_S$, where $d$ is the distance from ion to electrode and
$\delta_S$ is the skin depth of the electrode metal~\cite{henkel99}.
For $d \gg \delta_S$, the heating rate should scale as
$d^{-2}$~\cite{turchette00,henkel99,wineland75}. Heating due to
fluctuating surface patch potentials may scale more steeply.
Ref.~\cite{turchette00} indicates a scaling of~$d^{-4}$ if the size
of the patches is smaller than~$d$.

The geometry described above can easily be made smaller than most
current designs, and hence the trapped ions can be placed much
closer to the electrodes than in previous traps.    The spacing and
size of the planar electrodes may be changed as a function of
distance along the trap axis, therefore varying the ion-electrode
distance over a wide range on one chip. Tests of the scaling of
these heating rates is very important as traps become smaller. The
requirements for sympathetic cooling and constraints on ion motion
must be determined, and these depend crucially on the heating rate.

As in other small ion traps, there is a concern with accumulated
films of the to-be-trapped material on the trap electrodes.  There
has been some evidence that this material can lead to ion
motional-state heating at anomalously high
rates~\cite{mary,turchette00}. In light of the attainable proximity
of ions to the electrodes in the surface geometry, this may be even
more important in this case. Large-scale designs may require
separate loading and experimental zones as has been implemented in
previous designs~\cite{SPIE2005}.

Charging of trap or structural components due to electrons from
impact ionization may also lead to unpredictable heating or
ion-trapping behavior. A separate loading zone can alleviate this
problem if the experimental zone is sufficiently far away.
Alternately, photoionization may be employed if suitable radiation
sources are available; the usefulness of this technique is
ion-species-specific, however.

\section{Remarks}

We have explored the feasibility of some different linear ion-trap
electrode geometries in the context of multiplexed trapped-ion
quantum information processing.  Both four- or five-wire surface
geometries comprising metal deposited on an electrically insulating
substrate can maintain a trap axis above the chip and are promising
for scaling ion systems toward large-scale quantum information
processing. The fabrication of such trap chips is straightforward,
and allows for rapid turnaround in design and construction as well
as the inclusion of complicated electrode patterns without
additional fabrication effort. This design also allows for the
construction of traps much smaller than those currently in use, a
necessity for faster ion-qubit logic operations. Ion motional
heating, which may ultimately limit the size and speed of these ion
trap arrays, may also be explored in designs of this type.
Experiments whose goal is to trap atomic ions in this type of
electrode structure are currently underway. Planar traps of similar
design have been demonstrated recently for macroscopic charged
particles~\cite{ikepc}.  We have also suggested a modification of a
two-layer design to allow crosses at the intersection of two linear
traps to facilitate two-dimensional ion-trap arrays.

After submission of this manuscript, a publication suggesting a
planar Penning-trap design came to our
attention~\cite{planar_penning}.  In contrast to the RF Paul trap
designs discussed here, Penning traps use combined static electric
and magnetic fields to trap charged particles in three dimensions.

\nonumsection{Acknowledgements}

\noindent DJW thanks M.~Prentiss (Harvard University) for the
suggestion of planar electrode geometries. The authors thank
J.~Koelemeij and S.~Seidelin for helpful comments on the manuscript.
This work was supported by the U.S. National Security Agency (NSA)
and the Advanced Research and Development Activity (ARDA).  This
manuscript is work of the National Institute of Standard and
Technology and is not subject to U.S. copyright.

\nonumsection{References}

\newpage

\appendix{:  Analytic comparison of standard quadrupole and
four-wire surface geometries}


\label{app_analytic}

\noindent To compare the pseudopotential (in two dimensions) of the
four-wire surface trap to that of the standard quadrupole, complex
variables can be used (here we follow the method of~\cite{janik}).
In the following comparison we assume identical RF potentials
applied to electrode structures trapping particles of identical mass
and charge, and we assume a uniform static axial potential.  We
first calculate the pseudopotential of systems of line charges as a
limiting case of the typical rod or wire conductors and then discuss
finite-size effects.  In general, the results are directly
applicable to the case where the lateral electrode dimensions are
much smaller than their separation, but they give some qualitative
insight for the case proposed above, in which the spacings between
electrodes can be small.

\begin{figure}[tb]
\hbox{a) \hspace{0.46 \columnwidth} b)}
\centerline{\epsfig{file=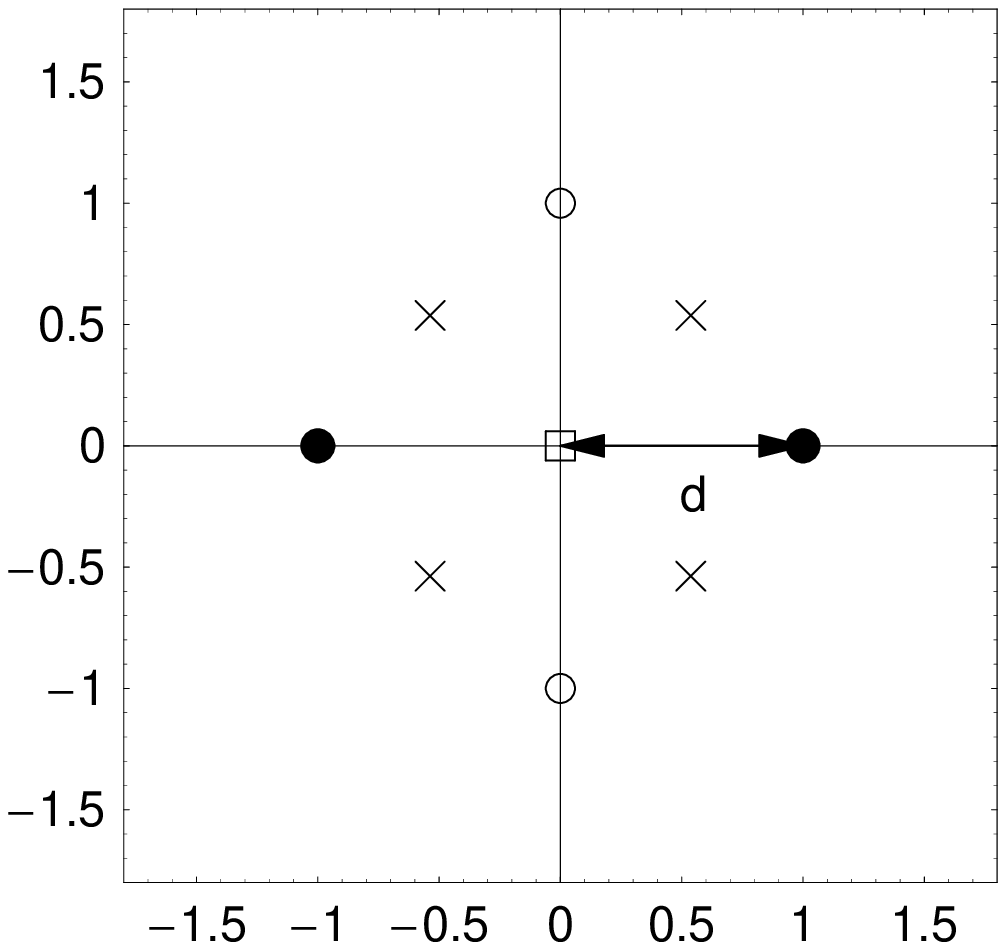, width=0.46 \columnwidth}
\hspace{0.04 \columnwidth} \epsfig{file=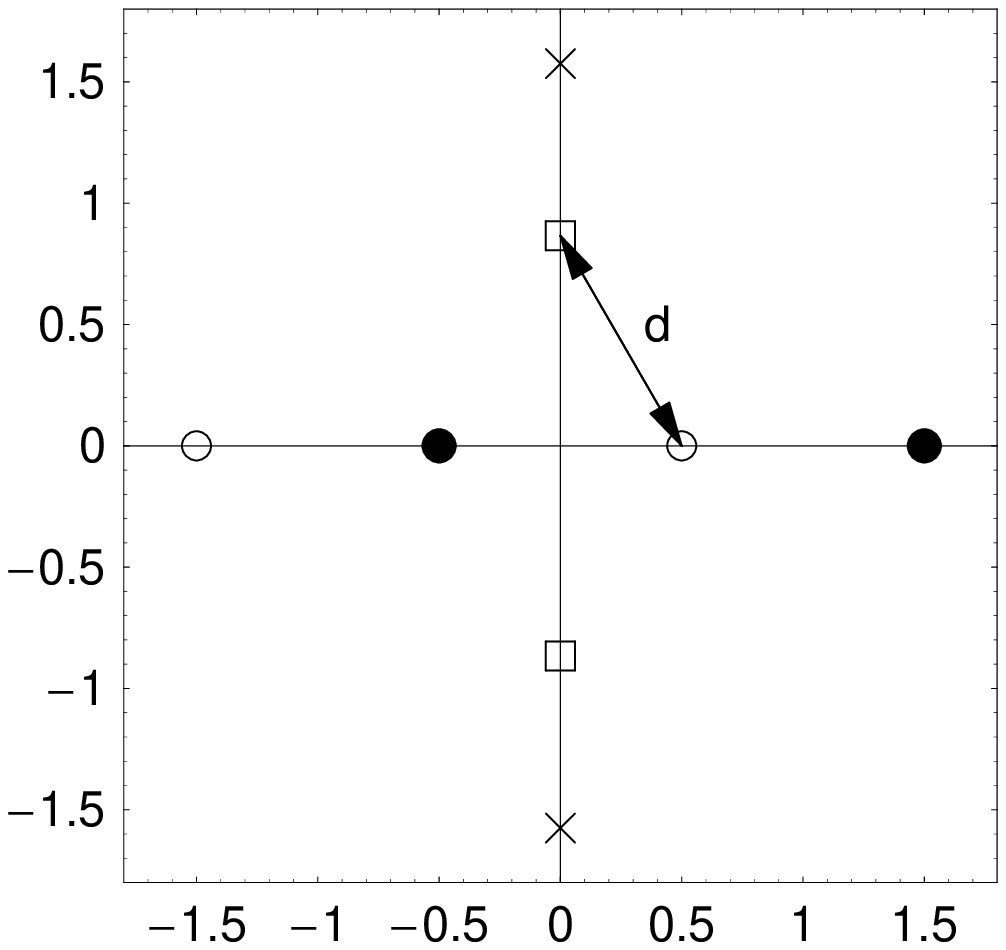, width=0.46
\columnwidth}} \vspace*{13pt} \fcaption{\label{analytic_layout}
Complex plane showing the locations of positive line charges
(`$\bullet$'), negative line charges (`$\circ$'), trap minima
(`$\square$'), and trap local maxima (`$\times$'). (a) Standard
quadrupole geometry.  (b) Four-wire surface electrode geometry.}
\end{figure}


The complex potential due to a positive line charge with charge per
unit length~$\lambda$ located at $z=x+iy=z_0$ is $w(z)=-(\lambda / 2
\pi \epsilon_0)\ln (z-z_0)$. The real part of this expression is the
electric potential.  The electric-field components $E_x$ and $E_y$
can be determined by means of the relation $dw/dz=-E_x + i E_y$. The
potential $w_{\rm q}(z)$ for a standard quadrupole trap would
consist of positive line charges at $z=\pm 1$ and negative line
charges at $z=\pm i$ (see Fig.~\ref{analytic_layout}a), which is
$w_{\rm q}(z)=\ln\left[ (z^2+1)/(z^2-1) \right]$ (here and below, we
let $\lambda / 2 \pi \epsilon_0=1$). Since the pseudopotential is
proportional to the square of the electric field, we calculate the
quantity

\begin{equation}
\Upsilon_{\rm q}(z) = \left|\frac{dw_{\rm q} }{ dz} \right|^2 =
\frac{16|z|^2 }{ |z^4-1|^2}.
\end{equation}

\noindent The function~$\Upsilon_{\rm q}(z)$ is not in standard
units but is suitable for comparisons between geometries.  The
pseudopotential of the quadrupole can be seen to have a minimum at
the origin, $\Upsilon_{\rm q}(0)=0$.  This is the trap axis, where
the distance $d$ from the axis to the nearest line charge is equal
to~1. The second spatial derivative of the pseudopotential (the
curvature) at this point can be found to be~${d^2 \Upsilon_q(z)
\over dz^2}|_{z=0}=32$. There are four local maxima at the points
$z=z_{\rm m}=\pm {1 \over \sqrt{2\sqrt{3}}}\pm i {1 \over
\sqrt{2\sqrt{3}}}$ with $\Upsilon_{\rm q}(z_{\rm m})=3\sqrt{3}$. The
minimum and local maxima are labeled in Fig.~\ref{analytic_layout}a.

Applying the same analysis to a four-wire surface geometry for
positive line charges at the points $z=-1/2$ and $z=3/2$ and
negative line charges at the points $z=-3/2$ and $z=1/2$ (see
Fig.~\ref{analytic_layout}b), we obtain the complex potential

\begin{equation}
w_{\rm s}(z)=\ln{ (z-{1 \over 2})(z+{3\over 2}) \over (z+{1 \over
2})(z-{3\over 2}) }.
\end{equation}

\noindent This gives a pseudopotential function (now switching
to~$x$ and~$y$)

\begin{equation}
\Upsilon_{\rm s}(x,y)={64\left(16\ x^4 + \left(3 - 4 y^2\right)^2 +
8 x^2 \left(3 + 4 y^2\right)\right)  \over \left(\left(1 - 2
x\right)^2 + 4 y^2\right) \left(\left(3 - 2 x\right)^2 + 4
y^2\right) \left(\left(1 + 2 x\right)^2 + 4 y^2\right)\left(\left(3
+ 2x\right)^2 + 4 y^2\right) }.
\end{equation}

\noindent  This is not particularly revealing, but differentiating,
one finds two pseudopotential minima $\Upsilon_{\rm
s}(0,\pm\sqrt{3}/2)\nobreak=\nobreak0$, which define the trap axes
above and below the surface, each a distance $d=1$ from the nearest
line charge. The pseudopotential curvature at this point is~8/3,
which is~1/12 of the curvature in the standard quadrupole case,
leading to radial secular frequencies~${1 \over
2\sqrt{3}}\approx0.29$~times as large as in the standard quadrupole.
There is a local maximum above the trap axis at
$y=y_m={\sqrt{3+4\sqrt{3}} \over 2}$ with $\Upsilon_{\rm
s}(0,y_m)={1 \over 7 + 4\sqrt{3}}$.  This gives a trap depth that is
lower by a factor of $3(12+7\sqrt{3})\approx 72$ than that of the
standard quadrupole trap.

To make an estimation more appropriate for finite-sized conductors,
we consider cylindrical, perfectly-conducting electrodes that
approximate equipotentials due to the line charges considered above.
In the surface configuration, more charge must be placed on the
inner two wires relative to the outer two wires in order to obtain
equipotentials of the same diameter around all four wires. These
equipotentials can be made the same size as those in the four-rod
geometry, and the near circularity of the equipotentials is such
that the vertical and horizontal diameters differ by only a few
percent. We numerically estimate that the inner two wires must have
approximately 1.35~times the charge of the outer two for an
electrode diameter of 0.2 (0.2~times the distance between electrode
centers). With this consideration, the above analysis can be
repeated, and we obtain trap axes at the locations (0,$\pm1.18$) and
local maxima at the locations (0,$\pm1.96$). The curvature at the
minima leads to a secular frequency approximately 0.16~times that in
a similar four-rod trap, and the trap depth is smaller by a factor
of approximately~200.

\appendix{:  Limits to miniaturization}

\label{app_scalelimit}

\noindent With any miniaturizable ion trap geometry, one may ask how
small the electrode structures can be made.  Here we calculate a
lower-length-scale limit, neglecting motional heating effects and
ignoring the requirement of optical access for laser cooling. We
assume stable trapping, and thus fix the normalized trap strength
parameter~$q_i$ (Eq.~\ref{q_eq}). For a particular ion species, this
leads to constraints on the applied RF potential amplitude and
frequency as~$R$ gets smaller (here we take $R$ to be the
approximate trap size). These constraints in turn restrict the trap
depth~$U_{\rm max}$, which must shrink as the trap size is reduced.
The following discussion concerns traps of the standard quadrupole
geometry (Fig.~\ref{quad}) for singly-ionized species. Parameters
for the other trap structures mentioned above can be derived from
the results of Sec.~\ref{sec_strength}.

If we demand that $q_i$ remain fixed as $R$ is reduced, then from
Eq.~\ref{q_eq},

\begin{equation}
{V_0 \over \Omega^2} = {m q_i \over 2 Q }R^2. \label{voveromega}
\end{equation}

\noindent  The potential amplitude must be made to scale
as~$V_0\propto R$ to ensure that the electric field amplitude at the
surfaces of the electrodes $E_0= V_0/R$ will not grow, leading to
breakdown, as the trap electrodes get smaller (here we have assumed
spherical electrode surfaces of radius~$R$, relevant for the
truncation of electrodes or for junctions in an array). From this
follows the condition that the RF frequency must scale
as~$\Omega\propto R^{-1/2}$.

The trap depth is equal to the value of the pseudopotential at the
location of the lowest nearby maximum; this location is displaced
from the trap minimum by approximately~$R$ in all geometries
studied, and will be proportional to~$R$ in any case. Thus, from
Eq.~\ref{pseud_quad}, the trap depth is

\begin{equation}
U_{\rm max}={\beta Q^2 V_0^2 \over 4 m \Omega^2 R^2}. \label{Umax}
\end{equation}

\noindent Here $\beta$ is a geometrical factor dependent on exact
electrode shape.  From Eqs.~\ref{voveromega} and~\ref{Umax}, we get

\begin{equation}
U_{\rm max}={\beta q_i Q \over 8} V_0={\beta q_i Q \over 8} E_0 R.
\end{equation}

\begin{figure}[tb]
\centerline{\epsfig{file=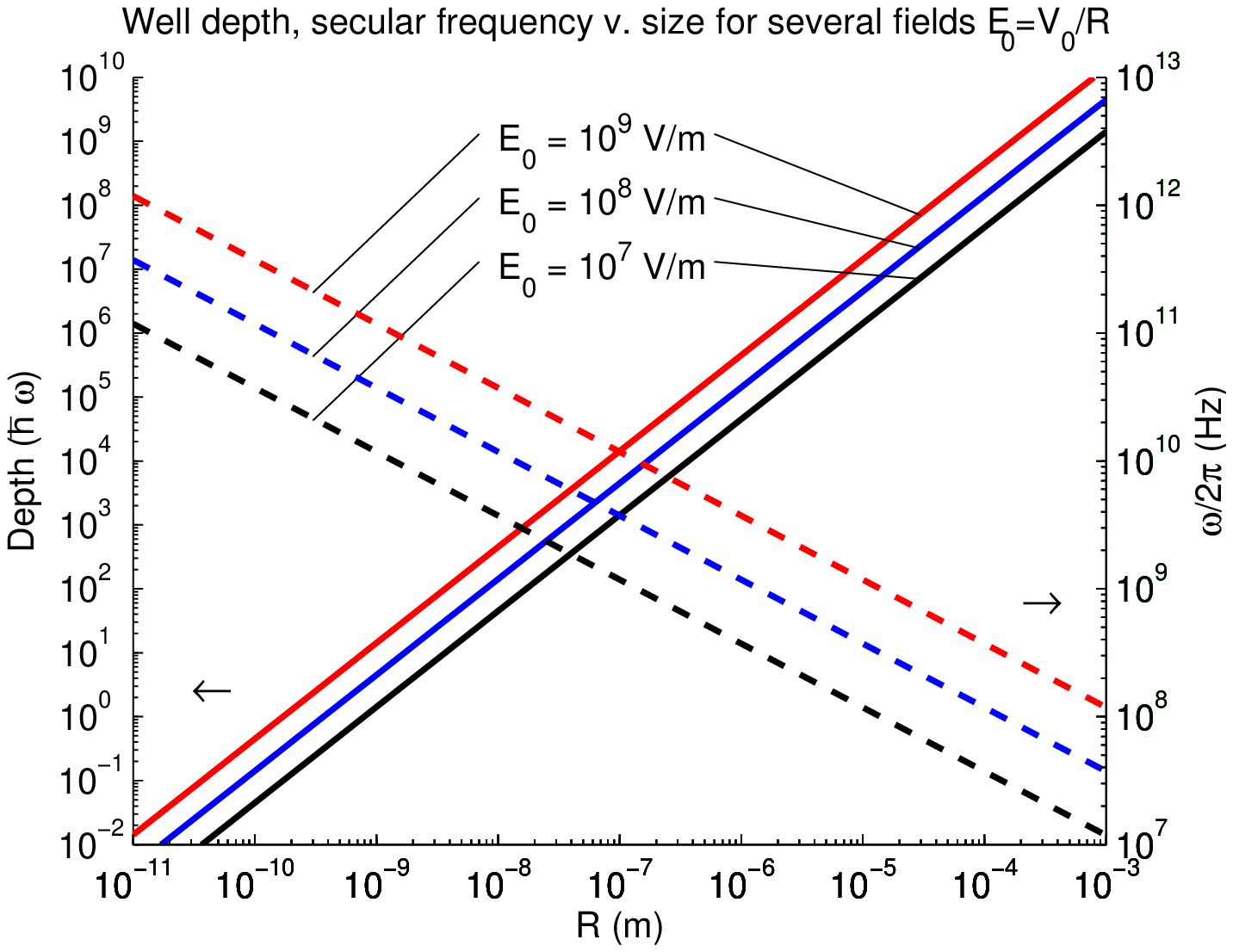, width=0.8 \columnwidth}}
\vspace*{13pt} \fcaption{\label{scaling_fig} Trap depth and secular
frequency as a function of trap size~$R$ for several maximum field
values~$E_0$.  The solid lines are depth, graphed on the left axis
in units of~$\hbar\omega$, and the dashed lines are secular
frequency~$\omega/2\pi$, graphed on the right axis.  For this graph,
$q_i=0.21$, $m=9$~u, and $\beta=0.34$.}
\end{figure}

The ion trap will cease to be useful when~$U_{\rm max}\approx \hbar
\omega_i$, that is, when there is only one trapped state in the
harmonic well.  The secular frequency~$\omega_i$ is approximately
equal to~$q_i \Omega / 2\sqrt{2}$, which for constant $q_i$ is

\begin{equation}
\omega_i = {1 \over 2}\sqrt{q_i Q E_0 \over m R}.
\end{equation}

\noindent  The ratio~$U_{\rm max} / \hbar \omega_i$ scales
as~$R^{3/2}$ and will be given by

\begin{equation}
{U_{\rm max} \over \hbar \omega_i}={\beta \over 4 \hbar}\sqrt{q_i Q
m E_0}\: R^{3/2}.
\end{equation}

Taking values of~$E_0=10^6$~V/m and $q_1=0.21$ for a beryllium ion
($m=9$~u) trapped in a four-rod quadrupole with $\beta=0.34$ (as in
the electrode size and configuration depicted in Fig.~\ref{quad}),
the trap will support at least one vibrational state (at a secular
frequency of approximately 2.9~GHz) for~$R\gtsim 1.7$~nm.  A more
realistic limit might be somewhat larger: for~$U_{\rm max} / \hbar
\omega_i=100$, $R\approx 37$~nm, for instance.  Current traps do not
operate near the breakdown electric field
($E_0\approx10^9$~V/m~\cite{home,gomer}), so smaller traps are
possible with higher maximum fields~$E_0$. Figure~\ref{scaling_fig}
shows the trap depth and secular frequency for different trap sizes
and several values of~$E_0$ up to $10^9$~V/m, where field emission
is possible. Operating slightly below this field, trapping
structures may shrink to $R\approx0.2$~nm before there are no
trapped states. At this point, the secular frequency would be a few
hundred gigahertz, with a required drive frequency of approximately
13.5~times larger. Traps with weaker intrinsic trapping potentials
compared to the four-rod quadrupole (such as all the other designs
suggested here) will have suitably larger minimum sizes using the
criteria presented here.

A recent proposal for scaling ion traps to the submicrometer scale
suggests using nanomechanical resonators (possibly even carbon
nanotubes) as trap electrodes and trapping an ion at a distance
of~$R=100$~nm from these electrodes~\cite{tian}.  The authors do not
consider trap depth in this work, and we point out that such a trap
would most likely need to be loaded from a larger structure, as the
depth may not be sufficient for initial ion capture.

\appendix{:  Substrate/insulator material properties}

\label{app_matprop}

\noindent Table~\ref{mat_table} lists values of pertinent properties
for a selection of materials both currently used and suggested for
use as insulators or substrates in the construction of
microfabricated ion traps.  These values are approximate and are
only intended to give an idea of the relative merit of particular
materials.

\vspace*{4pt}   
\begin{table}[bthp]
\tcaption{\label{mat_table} Substrate material properties.  Values
are given for readily available substrates at approximately room
temperature for frequencies in the RF range. In general, properties
of the crystalline materials vary in different directions or with
different substrate orientations; values are approximate. Properties
of semiconductors will vary greatly as a function of doping; values
are for low or intrinsic levels of doping. These values were
obtained from various published and unpublished sources.}
\centerline{\footnotesize\smalllineskip
\begin{tabular}{l l l l l l l}\\
\hline
Material & Thermal & Electric & Dielectric & Dissipation & Surface & Dielectric \\
{} & conductivity  & resistivity  & constant & factor & roughness  &strength\\
{} & (W/(m~K)) & ($\Omega$~cm) & & ($\tan\delta$) & (nm) & (kV/mm)\\
\hline
AlN               & \hfill 180  & $1\times10^{13}$ &\hfill 8.5 &$3\times10^{-4}$ &\hfill 50  &\hfill 16\\
Alumina (99.5~\%) & \hfill 30   & $1\times10^{14}$ &\hfill 9.8 &$1\times10^{-4}$ &\hfill 50  &\hfill 20\\
BN                & \hfill 28   & $1\times10^{13}$ &\hfill 4.1 &$5\times10^{-4}$ &\hfill 50  &\hfill 30\\
Diamond           & \hfill 2000 & $1\times10^{14}$ &\hfill 5.7 &$6\times10^{-4}$ & \hfill 50 &\hfill 1000\\
Fused silica      & \hfill 1    & $1\times10^{16}$ &\hfill 3.9 &$4\times10^{-4}$ & \hfill 2  &\hfill 40\\
GaAs              & \hfill 55   & $1\times10^{8}$  &\hfill 13  &$1\times10^{-3}$ &\hfill 25  &\hfill 40\\
Quartz            & \hfill 7    & $1\times10^{16}$ &\hfill 4.5 &$2\times10^{-4}$ &\hfill 2   & \hfill 80\\
Sapphire          & \hfill 45   & $1\times10^{14}$ &\hfill 11  &$1\times10^{-4}$ &\hfill 10  &\hfill 35\\
Si                & \hfill 150  & $1\times10^{5}$  &\hfill 12  &$5\times10^{-3}$ &\hfill 25  &\hfill 30\\
SiC               & \hfill 250  & $3\times10^{5}$  & \hfill 14 &$2\times10^{-1}$ &\hfill 50  & \hfill 300\\
\hline\\
\end{tabular}}
\end{table}

\end{document}